\documentclass[1p,times,review]{elsarticle}
\usepackage[english]{babel}
\usepackage{tensor}
\usepackage{graphicx,psfrag,subfigure}
\usepackage{amsmath}
\usepackage{amssymb}
\usepackage{amsfonts}
\usepackage{dcolumn}
\usepackage{bm}

\newcommand{\D}{\displaystyle}

\def\bra<#1|{\mathinner{\langle\,{#1}\,\vert}} 
\def\ket|#1>{\mathinner{\vert\,{#1}\,\rangle}} 
\def\red|#1|{\mathinner{\!\vert\,{#1}\,\vert\!}}
\def\braket<#1>{\mathinner{\langle\,{#1}\,\rangle}} 
\newcommand{\half}
	{ {\frac{1}{2}}}

\renewcommand{\vec}[1]{\ensuremath{{\mathbf{#1}}}}
\begin{document}
%%%%%%%%%%%%%%%%%%%%%%%%%%%%%%%%%%%%%%%%%%%%%%%%%%%%%%%%%%%%%%%%%%%%%%%%%%%%%%%
\title{Multiconfiguration electron density function for the ATSP2K-package}
\date{\today}

\author[vub]{A. Borgoo}
\address[vub]{%
Algemene Chemie, 
Vrije Universiteit Brussels, 
B--1050 Brussels, Belgium
}
\author[ulb]{O. Scharf}
\address[ulb]{%
Chimie quantique et Photophysique, 
Universit\'e Libre de Bruxelles, 
B--1050 Bruxelles, Belgium
}

\author[vil]{G. Gaigalas}
\address[vil]{%
Vilnius University Research Institute of Theoretical Physics and Astronomy, 
A. Go\v{s}tauto 12, LT-01108 Vilnius, Lithuania
}

\author[ulb]{M.~Godefroid}
\ead{mrgodef@ulb.ac.be}

%%%%%%%%%%%%%%%%%%%%%%%%%%%%%%%%%%%%%%%%%%%%%%%%%%%%%%%%%%%%%%%%%%%%%%%%%%%%%%%
%
%   ABSTRACT
%
%%%%%%%%%%%%%%%%%%%%%%%%%%%%%%%%%%%%%%%%%%%%%%%%%%%%%%%%%%%%%%%%%%%%%%%%%%%%%%%
\begin{abstract}
A new {\sc atsp2K} module is presented for evaluating the  electron density function of any multiconfiguration 
Hartree-Fock or configuration interaction wave function in the non relativistic or relativistic Breit-Pauli approximation. 
It is first stressed that the density function is not a priori  spherically symmetric 
in the general open shell case. Ways of building it as a spherical symmetric function are discussed, 
from which the radial electron density function emerges. This function is  written 
in second quantized coupled tensorial form for exploring the atomic spherical symmetry. 
The calculation of its expectation value is performed using the angular momentum theory in  orbital, 
spin, and quasispin spaces, adopting a generalized graphical technique. 
The natural orbitals are  evaluated from the diagonalization of the density matrix. 
\end{abstract}


\maketitle

{\bf  Program summary} \\

\begin{small}

\noindent {\em Title of program} {\em :} {\sc DENSITY}
\hspace{18pt}; {\em version number:} 1.00 \\[10pt]
{\em Catalogue identifier:} \\[10pt]
{\em Program obtainable from:} CPC Program Library, Queen's University

of Belfast, N. Ireland  \\[10pt]
{\em Computers:} HP XC Cluster Platform 4000  \\
{\em Installations:} VUB-ULB Computer Center \\ (http://www.vub.ac.be/BFUCC/hydra/about.html) \\[10pt]
{\em Operating systems or monitors under which the present version has been
tested:} HP XC System Software 3.2.1, which is a Linux distribution compatible with Red Hat Enterprise Advanced Server.
\\[10pt]
{\em Programming language used in the present version:} FORTRAN 90\\[10pt]
{\em RAM:} ??~MB or more \\[10pt]
{\em Peripherals used:} terminal, disk\\[10pt]
{\em No. of bits in a word:} 32 \\[10pt]
{\em No. of processors used:} 1\\[10pt]
{\em Has the code been vectorised or parallelized?:} no \\[10pt]
{\em No. of bytes in distributed program, including test data, etc.:}
??~bytes \\[10pt]
{\em Distribution format:} gzipped compressed tar file \\[10pt]
{\em CPC Program Library subprograms used:} libraries of {\sc atsp2K} \\[10pt]
{\em Nature of physical problem}\\
This program determines the atomic electronic density in the MCHF ($LS$) or Breit-Pauli ($LSJ$) approximation. It also evaluates the natural orbitals by diagonalizing the density matrix.
\\[10pt]
{\em Method of solution} \\
Building the density operator using second quantization - Spherical symmetry averaging - Evaluating the matrix elements of the one-body excitation operators in the 
configuration state function (CSF) space using the angular momentum theory in  orbital, 
spin, and quasispin spaces.
\\[10pt]
{\em Restrictions on the complexity of the problem}\\
Original restrictions from {\sc atsp2K} package, i.e. 
all orbitals within a wave function expansion are assumed to be orthonormal.
Configuration states are restricted to at most eight 
subshells in addition to the closed shells common to all configuration
states. The maximum size of the working arrays, related to the number of CSFs and active orbitals, 
is limited by the available memory and disk space.
\\[10pt]
{\em Typical running time }\\
The calculation of the electron density for a $n=9$ complete active space (CAS) MCHF wave function (271~733 CSFs - 45~orbitals) takes around 9 minutes on one AMD Opteron dual-core @ 2.4 GHz CPU.
\\[10pt]
{\em Unusual features of the program }\\
The programming style is
essentially F77 with extensions for the POINTER data type and
associated memory allocation. These have been available on workstations
for more than a decade, but their implementations are compiler dependent.
The present code has been installed and tested extensively using 
the Portland Group, pgf90, compiler. 
\\[10pt]
{\em References}

\begin{enumerate}
\item ``An MCHF atomic-structure package for large-scale calculation'', 
Charlotte Froese Fischer, Georgio Tachiev, Gediminas Gaigalas and Michel R. Godefroid,
Computer Physics Communications {\bf 176} (2007) 559-579

\end{enumerate}

\end{small}

\newpage

%%%%%%%%%%%%%%%%%%%%%%%%%%%%%%%%%%%%%%%%%%%%%%%%%%%%%%%%%%%%%%%%%%%%%%%%%%%%%%%

%%%%%%%%%%%%%%%%%%%%%%%%%%%%%%%%%%%%%%%%%%%%%%%%%%%%%%%%%%%%%%%%%%%%%%%%%%%%%%%
%
%   SECTION 1. INTRODUCTION
%
%%%%%%%%%%%%%%%%%%%%%%%%%%%%%%%%%%%%%%%%%%%%%%%%%%%%%%%%%%%%%%%%%%%%%%%%%%%%%%%

\section{Introduction}
\label{sec1}

In electronic structure theory there are several approaches to describe the
behavior of electrons in atoms and molecules. Most of them are based on the
wave nature of the particles, permitting the system to be described by wave
functions, as eigenstates of the Schr\"odinger equation. The Hohenberg--Kohn
(HK) \cite{HohKoh:64a} theorems on the other hand say that the electronic structure of a
system is completely determined by its ground state electron density
function. According to the HK theorems, the energy of any system can be
written as a functional of this density function. Based on these results,
within Density Function Theory (DFT), several methods have been developed to
describe atoms and molecules through their density function \cite{ParYan:89a}. 
The development of
density functionals which yield a system's energy has  become a major field
of research in Chemistry and Physics. Nowadays a lot of research is being done
to investigate how the electron density function describes the system. In
conceptual DFT for example, chemical reactivity indices are
defined, which indicate how a system behaves in a chemical reaction, by
considering perturbations to the electron density function. Although wave
function methods were well developed before, DFT is now the most widely
used electronic structure method. The wide spread use of DFT can be accounted to
the relative computational ease with which energies can be determined.  Where a
wave function describing an $N$-particle system involves the position- and spin-
coordinates of all electrons, a density function, describing the same system,
only depends on the coordinates of one particle. Following the work of McWeeny
\cite{Wee:92a}, one can try to extract physically essential features from the electron
density function. 

Some of the present authors have established the periodicity of the
atoms in Mendeleev's periodic Table by making an information theoretical analysis of the electron density 
functions as probability distributions \cite{Boretal:04a}. Another work quantifies the relativistic effects 
on the basis of a comparison of density functions calculated within the one-configuration Hartree--Fock 
and Dirac--Fock approximations \cite{Boretal:07a}. 

The present code is an extension of the Atomic Structure Package {\sc atsp2K}~\cite{Froetal:07a} 
for evaluating the atomic density function from non relativistic and relativistic (in the Breit-Pauli approximation) 
multiconfiguration {\em ab initio} wavefunctions of atomic systems, adopting an efficient approach for spin-angular integrations~\cite{Gaietal:97a,Gaietal:98a}.
It allows the investigation of correlation effects on the 
density function for any non-relativistic correlation model, and of relativistic effects in the Breit-Pauli approximation.

In quantum chemistry, the natural orbitals (NO) are known to provide a particularly efficient choice of single-particle states \cite{Dav:72a,Dav:76a}. Moreover, NO  give the most rapidly convergent approximation to the total wave function and are often used as a basis set for generating a better wave function in an iterative manner. In atomic physics, NO are rarely used, although they constitute the orbital basis of the reduced form of the MCHF expansions  for helium-like and nominal two-electron atomic systems \cite{Fro:77a}. It would be worthwhile to study their potential for more than two electrons in the search of efficient optimization strategies. The present code fills this gap by building the  natural orbitals through the diagonalization of the density matrix.

%%%%%%%%%%%%%%%%%%%%%%%%%%%%%%%%%%%%%%%%%%%%%%%%%%%%%%%%%%%%%%%%%%%%%%%%%%%%%%%
%
%   SECTION 2. THE DENSITY FUNCTION OPERATOR
%
%%%%%%%%%%%%%%%%%%%%%%%%%%%%%%%%%%%%%%%%%%%%%%%%%%%%%%%%%%%%%%%%%%%%%%%%%%%%%%%
\section{\label{sec2}On the symmetry of the density function}

In this section we start by formulating  the multiconfiguration wave function for a well defined atomic state,  
and we calculate the corresponding density function. From this calculation, we regain
the specific angular (non-spherical) dependence of the density function. We also present different ways for deriving a spherical electron density function.

\subsection{\label{secsec:2}The multiconfiguration many-electron wavefunction}
In the multiconfiguration approach, the $N$-electron wavefunction $\Psi_{\alpha L S M_L M_S}$
is a linear combination of $M$ configuration state functions (CSFs) $\Phi_{\alpha_i LS M_L M_S}$ which are eigenfunctions of the total angular momentum $L^2$, 
the spin momentum $S^2$ and their projections $L_z$ and $S_z$,   with eigenvalues 
$\hbar^2 L(L+1) $ , $  \hbar^2 S(S+1) $, $ \hbar M_L $ and $ \hbar M_S $, respectively 
%%%%%%%%%%%%%%%%%%%%%%
\begin{equation}
\label{MCHF_exp}
\Psi_{\alpha LSM_LM_S}(\vec{x}_1, \cdots \vec{x}_N) 
= 
\sum_{i=1}^M \; c_i \; \Phi (\alpha_i L S M_L M_S; \vec{x}_1, \cdots \vec{x}_N) \; .
\end{equation}
%%%%%%%%%%%%%%%%%%%%%%
The set of variables $\{ \vec{x}_j \}$ represent the electron's space
and spin coordinates  
$\vec{x}_j \equiv (\vec{r}_j ,\sigma_j)\equiv(r_j ,\vartheta_j ,\varphi_j ,\sigma_j)$. The individual {\sc CSF}s are built from a set of one-electron spin-orbitals,
%%%%%%%%%%%%%%%%%%%%%
\begin{equation}
\label{spin_orbital}
\psi_{nlm_{l}sm_{s}}(\vec{x}) 
= R_{nl}(r)Y_{lm_{l}}(\vartheta,\varphi)\chi_{s m_{s}}(\sigma)
=  \frac{1}{r}P_{nl}(r) Y_{lm_{l}}(\vartheta,\varphi)\chi_{s m_{s}}(\sigma) \; ,
\end{equation}
%%%%%%%%%%%%%%%%%%%%%%
where $R_{nl} (r)\equiv P_{nl}(r)/r$,  $Y_{lm_l}(\vartheta,\varphi)$ and $\chi_{s m_s}(\sigma)$ are the radial, the angular  and the spin parts of the one electron functions.
The mixing coefficients $\{ c_i \}$ and the radial functions $\{ R_{n_jl_j} (r) \}$ are solutions of the multiconfiguration Hartree-Fock 
method in the MCHF approach. For a given set of orbitals, the mixing coefficient may also be the solution of 
the configuration interaction (CI) problem. The relativistic corrections can be taken into account by diagonalizing the Breit-Pauli Hamiltonian 
\cite{Hibetal:91a} in the $LSJ$-coupled CSF basis to get the intermediate coupling eigenvectors
\begin{equation}
\label{MCHF_BP_exp}
\Psi_{\alpha J M}(\vec{x}_1, \cdots \vec{x}_N) 
= 
\sum_{i=1}^{M'} \; a_i \; \Phi (\alpha_i L_i S_i  J M ; \vec{x}_1, \cdots \vec{x}_N) \; .
\end{equation}

\subsection{\label{secsec:2} The non-spherical density function }

The so-called ``generalized density function"  \cite{Wee:92a} or the ``first order reduced density matrix"  
\cite{Heletal:00a} is  a special case of the reduced density matrix  \cite{Dav:76a,Wee:92a} 
\begin{equation}
\label{gamma_1}
\gamma_1(\vec{x}_1,\vec{x'}_1) = 
N \int \Psi(\vec{x}_1,\vec{x}_2, \ldots , \vec{x}_N)  \; \Psi^*(\vec{x'}_1,\vec{x}_2, \ldots , \vec{x}_N) \; d\vec{x}_2 \ldots d \vec{x}_N \; ,
\end{equation}
where $\Psi(\vec{x}_1,\vec{x}_2, \ldots , \vec{x}_N)$ is the total wave function of an $N$ electron system and 
$\Psi^* (\vec{x}_1,\vec{x}_2, \ldots , \vec{x}_N)$ is 
its complex conjugate. The spin-less total electron density function $\rho(\vec{r})$ is defined as the first 
order reduced density matrix, integrated over the spin and evaluated for $\vec{x}_1 = \vec{x'}_1$
\begin{equation}
\rho(\vec{r}_1) = \int \gamma_1(\vec{x}_1,\vec{x}_1) d\sigma_1. \label{density}
\end{equation}
This electron density function is normalized to the number of electrons of the system
\begin{equation}
\int \rho(\vec{r})  \; d \vec{r} =  \int \rho(\vec{r}) \;  r^2  \sin\vartheta  dr d\vartheta d\varphi = N \; .
\end{equation}
As discussed in \cite{Heletal:00a}, the single particle density function can be calculated by evaluating the expectation value of the 
$ \delta (\vec{r}) $ operator, 
\begin{equation}
\label{xpectation_value}
\rho(\vec{r}) = 
\int \Psi(\vec{x}_1,\vec{x}_2, \ldots , \vec{x}_N) \;
\delta (\vec{r}) \;
\Psi^*(\vec{x}_1,\vec{x}_2, \ldots , \vec{x}_N) \; d\vec{x}_1 d\vec{x}_2 \ldots d \vec{x}_N \; ,
\end{equation}
where  $ \delta (\vec{r}) $ probes the presence of electrons at a particular point in space and can be written as the one-electron first-quantization operator
\begin{equation} 
\label{firstquant}
\delta (\vec{r}) = \sum_{i=1}^{N}\delta (\vec{r} - \vec{r}_i) \, .
\end{equation}
Expressing each $\delta (\vec{r} - \vec{r}_i)$ term in spherical coordinates~\cite{Cohetal:94a}
\begin{equation} 
\label{delta_sc}
\delta (\vec{r} - \vec{r}_i) =
\frac{1}{r^2 \sin \vartheta} \; \delta (r - r_i) \; \delta (\vartheta - \vartheta_i) \; \delta (\varphi - \varphi_i)  \, ,
\end{equation}
and introducing the closure relation
\begin{equation} 
\label{exp_sh}
\sum_{lm} Y_{lm} (\vartheta, \varphi) Y_{lm} ^\ast (\vartheta ', \varphi ') =
\delta (\cos \vartheta - \cos \vartheta ' ) \; \delta (\varphi - \varphi ')  \, ,
\end{equation}
the operator~(\ref{firstquant}) becomes
\begin{equation} 
\label{firstquant2}
\delta (\vec{r}) = \sum_{i=1}^{N}\delta (\vec{r} - \vec{r}_i) 
= \frac{1}{r^2 } \sum_{i=1}^{N} \left[ \delta (r - r_i) \; \sum_{lm} Y_{lm} (\vartheta, \varphi) Y_{lm} ^\ast (\vartheta_i, \varphi_i) \right]
\; .
\end{equation}
The exact spin-less total electron density function (\ref{xpectation_value}) evaluated for an eigenstate with well-defined quantum numbers $(L S M_L M_S)$, is
\begin{equation} 
\label{exact_density}
\rho(\vec{r}) ^{LSM_LM_S} 
%& = & \int \Psi_{\alpha LSM_LM_S}(\vec{x}_1, \cdots \vec{x}_N) \; \delta (\vec{r}) \; \Psi^{\ast}_{\alpha LSM_LM_S}(\vec{x}_1, \cdots \vec{x}_N) 
% \; d\vec{x}_1 d\vec{x}_2 \ldots d \vec{x}_N  \nonumber \\ 
% & = & \int \Psi_{\alpha LSM_LM_S}(\vec{x}_1, \cdots \vec{x}_N) \; \frac{1}{r^2 } \sum_{i=1}^{N} 
% \left[ \delta (r - r_i) \; \sum_{lm} Y_{lm} (\vartheta, \varphi) Y_{lm} ^\ast (\vartheta_i, 
% \varphi_i) \right] \; \Psi^{\ast}_{\alpha LSM_LM_S}(\vec{x}_1, \cdots \vec{x}_N) 
%  \; d\vec{x}_1 d\vec{x}_2 \ldots d \vec{x}_N \nonumber  \\ 
%& = & \langle \Psi_{\alpha LSM_LM_S} \vert   \frac{1}{r^2 } \; \sum_{i=1}^{N} \left[ \delta (r - r_i) \; \sum_{lm} Y_{lm} (\vartheta, \varphi) Y_{lm} ^\ast (\vartheta_i, \varphi_i) \right] \vert \Psi_{\alpha LSM_LM_S} \rangle
% \nonumber  \\ 
=   \sum_{lm} Y_{lm} (\vartheta, \varphi) \frac{1}{r^2 } \; 
\langle \Psi_{\alpha LSM_LM_S} \vert     
\sum_{i=1}^{N}  \delta (r - r_i) \;   Y_{lm} ^\ast (\vartheta_i, \varphi_i)  
\vert \Psi_{\alpha LSM_LM_S} \rangle
\end{equation} 
It is important to realize that the spherical harmonic components are limited to the 
$l$-even contributions, since the bra and ket states have the same parity $\pi = (-1)^{\sum_i{l_i}}$. 
Applying the Wigner-Eckart theorem \cite{Cow:81a} gives
\begin{eqnarray} 
\label{exact_density_2}
\rho(\vec{r}) ^{LSM_LM_S}
& = &  \sum_{l_{\mbox{\tiny even}}=0}^{2L} Y_{l \, 0} (\vartheta, \varphi) \frac{1}{r^2 } \; 
(-1)^{L - M_L} 
\left( \begin{array}{ccc} L & l & L \\ -M_L  & 0 & M_L \end{array} \right)
\langle \Psi_{\alpha LS } \Vert     \sum_{i=1}^{N}  \delta (r - r_i) \;   Y_{l} ^\ast (\vartheta_i, \varphi_i)  \Vert \Psi_{\alpha LS} \rangle      \nonumber  \\ 
& = & \sum_{l=0}^{L} \rho (r) ^{LSM_L M_S}_{2l} Y_{2l \; 0} (\vartheta, \varphi)
\end{eqnarray}
% & = & \sum_{l=0}^L \rho_{2l}(r) Y^0_{2l} (\vartheta, \varphi)
where 
\begin{equation}
\label{rho_2l}
\rho (r) ^{LSM_L M_S}_{2l} = 
\frac{1}{r^2 } \; 
(-1)^{L - M_L} 
\left( \begin{array}{ccc} L & 2l & L \\ -M_L  & 0 & M_L \end{array} \right)
\langle \Psi_{\alpha LS } \Vert     \sum_{i=1}^{N}  \delta (r - r_i) \;   Y_{2l} ^\ast (\vartheta_i, \varphi_i) \Vert \Psi_{\alpha LS} \rangle \; .
\end{equation}
This result\footnote{The same result can be obtained by reducing the many-electron reduced matrix element 
as a sum over one-electron reduced matrix elements as done in \cite{Rob:73a}.}
recovers Fertig and Kohn's analysis~\cite{FerKoh:00a} for the density corresponding to a 
well-defined $(LSM_LM_S)$ eigenstate of the Schr\"{o}dinger equation. In this paper, the authors
observed that the self-consistent field densities obtained 
via the Hartree and Hartree-Fock methods generally violate the specific finite spherical harmonic 
content of $\rho(\vec{r}) ^{LSM_LM_S}$.
%~(\ref{exact_density_2})
They also mention that this exact form can be obtained by 
spherically averaging the effective potential, yielding single-particle states with good angular 
momentum quantum numbers. 
The atomic  structure software package {\sc atsp2K}~\cite{Froetal:07a} applies this approach, as was done in the original 
atomic Hartree-Fock theory \cite{Sla:30a,Har:57a,Fro:77a}. This implies two things: 
i) the density function $\rho (\vec{r})^{LSM_LM_S}$ calculated from any multiconfiguration wave function 
of the form (\ref{MCHF_exp}), is not {\em a priori} spherically symmetric, 
ii) this density function will contain all spherical harmonic components (up to $2L$) as long as 
the one-electron orbital active set spanning the configuration space is $l$-rich enough. 

The density function can also be expressed in second quantization \cite{Wee:92a}. Introducing the notation 
$ q \equiv n_q l_q m_{l_q} m_{s_q} $ for spin-orbitals, expression~(\ref{gamma_1}) becomes 
\begin{equation}
\gamma_1(\vec{x}_1,\vec{x'}_1) = \sum_{pq} D_{pq}  \; \psi^*_p(\vec{x'}_1) \psi_q(\vec{x}_1) \label{gamma} \; ,
\end{equation}
where $D_{pq}$ are elements of the density matrix which are given by
\begin{equation}
D_{pq} \equiv 
\braket< \Psi |a^\dag_p a_q|\Psi> \; .
\end{equation}
The sum in eq.~(\ref{gamma}) runs over all possible pairs of quartets of quantum numbers 
$p$ and $q$. The spin-less density function~(\ref{density}) calculated from 
$ \rho(\vec{r}) =  \braket  < \Psi \vert \hat{\delta}(\vec{r})  \vert \Psi >$, 
using the second quantized form of the operator (\ref{delta_sc}) 
\begin{eqnarray}
\label{secont_quant_1}
\hat{\delta}(\vec{r}) & \equiv & \sum_{pq} a^\dag_p a_q \; \delta_{m_{s_p},m_{s_q}} 
\langle \psi_p(\vec{r'}) 
\vert   \frac{1}{r^2 \sin \vartheta} \; \delta (r - r') \; \delta (\vartheta - \vartheta') \; \delta (\varphi - \varphi')  \vert 
\psi_q(\vec{r'}) \rangle   
\nonumber \\
& = & \sum_{pq} a^\dag_p a_q \; \delta_{m_{s_p},m_{s_q}} R^\ast_{n_p l_p}(r) Y^\ast_{l_p m_{l_p}}(\vartheta,\varphi) 
R_{n_q l_q}(r)    Y_{l_q m_{l_q}}(\vartheta,\varphi) \; ,
\end{eqnarray}
yields
\begin{equation}
\label{rho_r}
\rho(\vec{r}) =  
%\sum_{pq}  \braket  < \Psi \vert  \; a^\dag_p a_q   \vert \Psi >  \;  \delta_{m_{s_p},m_{s_q}} \;
%\psi_p^*(\vec{r}) \psi_q(\vec{r}) = 
\sum_{pq}    D_{pq} \; \delta_{m_{s_p},m_{s_q}} \; 
R^\ast_{n_p l_p}(r)Y^*_{l_p m_{l_p}}(\vartheta,\varphi) R_{n_q l_q}(r)Y_{l_q m_{l_q}}(\vartheta,\varphi) \; .
\end{equation}

To illustrate the spherical harmonics content of the density in the Hartree-Fock approximation, 
consider the atomic term $ 1s^2 2p^2 (\; ^3P) 3d \; ^4F$ for which the $(M_L,M_S) = (+3,+3/2)$ subspace reduces to a single Slater determinant 
\begin{equation}
\Psi_{\alpha L S M_L M_S} = 
\Phi (1s^2 2p^2 (\; ^3P) 3d \; ^4F_{+3,+3/2} )   =
|1s \overline{1s} 2p_{+1} 2p_{0} 3d_{+2} | \; .
\end{equation}
When evaluating (\ref{rho_r}), all non-zero $D_{pq}$-values appear on the diagonal ($p=q$), yielding 
\begin{equation}
\label{not_spher}
\rho(\vec{r})^{^4F_{+3,+3/2}} =
|\psi_{1s}(\vec{r}) |^2 + |\psi_{\overline{1s}} (\vec{r}) |^2 + |\psi_{2p_{+1}} (\vec{r})|^2 + |\psi_{2p_0} (\vec{r})|^2 + |\psi_{3d_{+2}} (\vec{r})|^2
\; .
\end{equation}
This density has a clear {\em non}-spherical angular dependence.
However, referring to~\cite{Varetal:88a},
\begin{equation}
W^{\parallel}_{JM} (\vartheta) \equiv \vert Y_{JM}(\vartheta,\varphi) \vert^2 
= \sum_{n=0}^J b_n (J,M) \; P_{2n} (\cos \vartheta )
= \sum_{n=0}^J b'_n (J,M) \; Y_{2n \; 0} (\vartheta,\varphi)
\end{equation}
one recovers the even Legendre polynomial content of the density, although not 
reaching the  $(2L = 6)$ limit $Y_{6 \; 0}(\vartheta, \varphi) $ of 
the exact density (\ref{exact_density_2}).  However this limit will be attained  
when extending the one-electron orbital active set to higher angular momentum values for building 
a correlated wave function. 

Mixed contributions $(p \neq q)$ may appear in (\ref{rho_r}) through off-diagonal matrix elements 
in the CSF basis. For example, the interaction of $\Phi (1s^2 2p^2 (\; ^3P) 3d \; ^4F_{+3,+3/2} )$
with the angular correlation component $\Phi (1s^2 2p 3d (^3F) 4f \; ^4F_{+3,+3/2} ) $ , a single electron
excitation $2p \rightarrow 4f$, gives rise to $Y^\ast_{10} Y^{}_{30} $ and $Y^\ast_{1+1} Y^{}_{3+1} $
contributions. But these contributions are also limited to even Legendre polynomials, as appearing in
equation (\ref{exact_density_2}). Indeed, starting from the Clebsch-Gordan series~\cite{Varetal:88a}
\begin{equation}
Y_{l_1 m_1} (\vartheta,\varphi) Y_{l_2 m_2} (\vartheta,\varphi) 
%= \sum_{l=\vert l_1 - l_2 \vert}^{l_1+l_2}
%\sum_{m=-l}^l 
%\left[ \frac{(2l_1+1)(2l_2+1)}{4 \pi (2l+1)} \right] ^{1/2}
%\langle l_1 l_2 0 0 \vert l 0 \rangle 
%\langle l_1 l_2 m_1 m_2 \vert l m \rangle Y_{lm} (\vartheta,\varphi)
= \sum_{l=\vert l_1 - l_2 \vert}^{l_1+l_2}
\sum_{m=-l}^l 
\left[ \frac{(2l_1+1)(2l_2+1)(2l+1)}{4 \pi } \right] ^{1/2}
(-1)^m
\left( \begin{array}{ccc} l_1 & l_2 & l \\ 0  & 0 & 0 \end{array} \right)
\left( \begin{array}{ccc} l_1 & l_2 & l \\ m_1  & m_2 & -m \end{array} \right)
Y_{lm} (\vartheta,\varphi)
\end{equation}
and using
\begin{equation}
Y_{l -m } (\vartheta,\varphi) =(-1)^{m} Y^\ast_{lm} (\vartheta,\varphi) \; ,
\end{equation}
one finds that any contribution of the type 
$Y^\ast_{l_1 q} Y^{}_{l_2 q} $
arising from a single electron excitation $\vert l_1 q \rangle \rightarrow \vert l_2 q \rangle$ preserving the parity, ie. $(-1)^{l_1} = (-1)^{l_2}$, takes the form
\begin{equation}
Y^\ast_{l_1 q} (\vartheta,\varphi) Y_{l_2 q} (\vartheta,\varphi) 
= (-1)^q 
\sum_{l_{\mbox{\tiny even}}= \vert l_1 - l_2 \vert}^{l_1+l_2}
\left[ \frac{(2l_1+1)(2l_2+1)(2l+1)}{4 \pi } \right] ^{1/2}
\left( \begin{array}{ccc} l_1 & l_2 & l \\ 0  & 0 & 0 \end{array} \right)
\left( \begin{array}{ccc} l_1 & l_2 & l \\ -q  & +q & 0 \end{array} \right)
Y_{l0} (\vartheta,\varphi) \; .
\end{equation}

\noindent At this stage, we would like to stress that in an MCHF calculation
the  density never contains -- what Fertig and Kohn~\cite{FerKoh:00a} called -- ``offending'' spherical harmonic 
components, whatever the maximum $l$-value  of the orbital active space.

\subsection{\label{secsec:3}The spherical density function }

A {\em spherically} symmetric density function can be obtained for an arbitrary  CSF 
$\Phi_{\alpha LS M_L M_S}$ by averaging the  $(2L+1)(2S+1)$
magnetic components of the spin-less density function 
\begin{equation}
\label{av}
\rho(\vec{r}) ^{LS} \equiv \frac{1}{(2L+1)(2S+1)} \; \sum_{M_L M_S} \rho(\vec{r}) ^{LSM_LM_S} \; ,
\end{equation}
where $ \rho(\vec{r}) ^{LSM_LM_S} $ is constructed according to eq.~(\ref{rho_r})
\begin{equation}
\label{rho_r_LS}
\rho(\vec{r}) ^{LSM_LM_S} = \sum_{pq} \braket< \Phi_{\alpha LS M_L M_S} |a^\dag_{p} a_{q}|
\Phi_{\alpha LS M_L M_S} >  \;  \delta_{m_{s_p},m_{s_q}} \; \psi^*_p(\vec{r}) \psi_q(\vec{r}) \; .
\end{equation}

Applying equations~(\ref{av}) and (\ref{rho_r_LS})
for the atomic term $ 1s^2 2p^2 (\; ^3P) 3d \; ^4F$ considered in the previous section, 
we simply get
\begin{equation}
\rho(\vec{r}) ^{^4F} = \frac{1}{4 \pi r^2}  \left\{
2 P^2_{1s}(r)  + 2 P^2_{2p}(r) + P^2_{3d}(r) \right\} \; .
\end{equation}
which is, {\em in contrast} to eq.~(\ref{not_spher}), obviously spherically symmetric.
The sum over $(M_L,M_S)$ performed in (\ref{av}) 
guarantees, for any $nl$-subshell, the presence of all necessary components $ \{ Y_{lm_l} \; \vert \; m_l = -l, \ldots +l \} $ 
with the  same weight factor,   which permits the application of Uns\"{o}ld's theorem \cite{Uns:27a}
\begin{equation}
\label{Unsold}
\sum_{m_l=-l}^{+l} \; \vert Y_{lm_l} (\vartheta, \varphi) \vert ^2 = \frac{2l+1}{4 \pi} 
\end{equation}
and yields the spherical symmetry.  This result  is valid for any single CSF 
\begin{equation}
\label{gen}
\rho(\vec{r}) ^{LS} = \frac{1}{4 \pi r^2}  \sum_{nl} q_{nl}  P^2_{nl}(r)  \; ,
\end{equation}
where $q_{nl}$ is the occupation number of $n l$-subshell. Its sphericity explicitly appears by rewriting (\ref{gen}) as
\begin{equation}
\label{rho_r_vec}
\rho(\vec{r})  = \rho(r) \; \vert Y_{00} (\vartheta, \varphi) \vert ^2 
= \frac{D(r)}{r^2} \; \vert Y_{00} (\vartheta, \varphi) \vert ^2   \; ,
\end{equation}
with
\begin{equation}
\label{rho_r_scal}
\rho(r) \equiv \frac{1}{r^2} \sum_{nl} q_{nl}   P^2_{nl}(r) \; ,
\end{equation}
and 
\begin{equation}
\label{D_r}
D(r) \equiv r^2 \rho(r) = \sum_{nl} q_{nl}   P^2_{nl}(r)  = 
\sum_{nl} q_{nl} \; r^2  R^2_{nl}(r) \; .
\end{equation}\\
The {\em radial distribution} function $D(r)$ represents the probability of finding an electron 
between the distances $r$ and $r + dr$ from the nucleus, regardless of direction\footnote{Note that, although denoted as $D$, 
this function (evaluated at the $r=0$) is {\it not} the so-called  ``modified electron density'' used in the context of isotope shifts~\cite{Auf:82a}. 
The latter is indeed $\rho(0) = 4 \pi \rho({\bf 0}) $.}.
This radial density function reveals the 
atomic shell structure when plotted as function of $r$. Its integration over $r$ gives the total number of electrons of the system
\begin{equation}
\label{integral_density}
\int_{0}^\infty D(r)  \; d r = \int_{0}^\infty r^2 \rho(r)  \; d r =  \sum_{nl} q_{nl} = N \; .
\end{equation}

Where above the spherical symmetry of the average density (\ref{av}) is  demonstrated for a single CSF thanks to Uns\"{o}ld's theorem, it can be
demonstrated in the general case by combining (\ref{av}), (\ref{exact_density_2}) 
and the 3-$j$ sum rule \cite{Cow:81a}
\begin{equation}
\sum_{M_L} (-1)^{L - M_L} 
\left( \begin{array}{ccc} L & k & L \\ -M_L  & 0 & M_L \end{array} 
 \right) =  (2k + 1)^{1/2} \; \delta_{k,0}
\end{equation}
for each $k = 2l$ contribution~(\ref{rho_2l}). However, the radial density $\rho(r)$ will be more complicated than (\ref{rho_r_scal}), 
involving mixed contributions of the type 
$P_{n'l}(r) P_{nl}(r) = r^2 R_{n'l}(r) R_{nl}(r)$, as developed below.

Instead of obtaining a spherically symmetric density function by averaging the magnetic components $ \rho(\vec{r}) ^{LSM_LM_S}$ through
eq.~({\ref{av}), one can build a radial density operator associated to the function (\ref{D_r})  
which is spin- and angular-independent, i.e. independent of the spin~($\sigma$) and 
angular ($\vartheta, \varphi$) variables. Adopting the methodology used by 
Helgaker {\em et al} \cite{Heletal:00a} for defining the spin-less density operator, 
we write a general first quantization spin-free {\it radial} operator 
\begin{equation}
\label{oneparticle_FQ}
f = \sum_{i=1}^N f(r_i)
\end{equation}
in second quantization as
\begin{equation}
\label{oneparticle_SQ}
\hat{f} = \sum_{pq} f_{pq}  \; a_p^\dag a_q   \; ,
\end{equation}
where $f_{pq}$ is the one-electron integral
\begin{equation}
f_{pq} = \int \psi^*_p(\vec{x}) f(r) \psi_q(\vec{x}) r^2 \sin \vartheta  dr d\vartheta d\varphi d\sigma
\; .
\end{equation}
Applying this formalism to the radial density operator
\begin{equation} 
\label{first_quant_delta_r}
\delta (r) \equiv \sum_{i=1}^{N}\delta (r - r_i) \, ,
\end{equation}
and  using the spin-orbital factorization (\ref{spin_orbital}) for both $p$ and $q$ quartets, we obtain the
second quantization form 
\begin{equation}
\label{second_quant_delta_r}
\hat{\delta}(r) = \sum_{pq} d_{pq}(r)  \, a_p^\dag a_q \; ,
\end{equation}
with
\begin{equation} 
\label{delta_r_pq}
d_{pq}(r) =
\delta_{l_p l_q} \; \delta_{m_{l_p} m_{l_q} } \; \delta_{m_{s_p} m_{s_q}}  \; R^*_{n_pl_p} (r) R_{n_ql_q} (r) r^2
\; ,
\end{equation}
where the Kronecker delta arises from the orthonormality property of the spherical harmonics and spin functions.
With real radial one-electron functions, the operator (\ref{second_quant_delta_r}) becomes
\begin{equation}
\label{second_quant_delta_r_2}
\hat{\delta}(r) = 
\sum_{n', l', m_l', m_s', n, l, m_l, m_s, } 
\delta_{l' l} \; \delta_{m'_l m_l} \; \delta_{m'_s m_s} \; a^\dag_{n'l'm'_lm'_s} a_{nlm_lm_s} \;
R_{n'l'} (r) R_{nl} (r) r^2 
\end{equation}
\begin{equation}
= \sum_{n',n}  \sum_{l, m_l, m_s}  \; a^\dag_{n'lm_lm_s} a_{nlm_lm_s} \; R_{n'l} (r) R_{nl} (r) r^2  \; .
\end{equation}
Its expectation value  provides the radial density function 
$D(r) = r^2 \rho(r) = 4 \pi r^2 \rho( \vec{r})$ defined by (\ref{rho_r_vec}) 
and (\ref{D_r}).

Building the coupled tensor of ranks $(00)$   from the $[2(2l+1)]$ components of the 
creation and annihilation operators \cite{Jud:67a}
\begin{equation}
\label{coupled_tensor_00}
\left(
{\bf a}^\dag_{n'l} {\bf a}_{nl} \right) ^{(00)}_{00}
= -\frac{1}{\sqrt{2(2l+1)}} \; \sum_{m_l m_s} 
a^\dag_{n'lm_lm_s} a_{nlm_lm_s}   \; ,
\end{equation}
the operator (\ref{second_quant_delta_r_2}) becomes
\begin{equation}
\label{second_quant_delta_r_3}
\hat{\delta}(r) = - \sum_l \sqrt{2(2l+1)} 
\sum_{n',n}  \; \left( {\bf a}^\dag_{n'l} {\bf a}_{nl} \right) ^{(00)}_{00} 
\; R_{n'l} (r) R_{nl} (r) r^2  \; .
\end{equation}
The expectation value of this operator provides the spherical density function for any atomic state.
Note that, in contrast to (\ref{rho_r_LS}), the tensorial ranks (00) garantee the diagonal character in $L, S, M_L $ and $M_S$, 
thanks to Wigner-Eckart theorem 
\begin{equation}
\label{WE_00}
\langle \alpha L S M_L  M_S \vert T^{(00)}_{00} \vert \alpha' L' S' M_L'  M_S' \rangle
= (-1)^{L + S - M_L - M_S} 
\left( \begin{array}{ccc} L & 0 & L' \\ -M_L  & 0 & M_L' \end{array}    \right)
  \left( \begin{array}{ccc} S & 0 & S' \\ -M_S  & 0 & M_S' \end{array}    \right) 
  \langle \alpha L S \Vert T^{(00)} \Vert \alpha' L'  S'  \rangle\; .
\end{equation}
Moreover, the $M_L / M_S$ independence emerges from the special $3j$-symbol
% Cowan (5.11)
\begin{equation}
\label{special_3j}
\left( \begin{array}{ccc} j & 0 & j' \\ -m_j  & 0 & m_j' \end{array}    \right)
= (-1)^{j-m} (2j + 1)^{-1/2} \delta_{jj'} \delta_{m{_j} m_{j}'} \; .
\end{equation}
In other words, where the non-spherical components are washed out by the averaging process (\ref{av}), 
they simply do not exist and will never appear for the 
density calculated from (\ref{second_quant_delta_r_3}), for any $(M_L, M_S)$ magnetic component.

% from GG

%According to the second-quantization (see (A5) in \cite{Heletal:00a} density 
The radial distribution function $D(r) \equiv r^2 \rho(r)$ can be calculated from the expectation value of the operator (\ref{second_quant_delta_r_3}),
using the wave function (\ref{MCHF_exp}) or (\ref{MCHF_BP_exp}). In the most general case (expansion (\ref{MCHF_BP_exp})), using the $(LS)J$-coupled form of the
excitation operator, 
\begin{equation}
%\label{?}
\left( {\bf a}^\dag_{n'l} {\bf a}_{nl} \right) ^{(00)0}_{0}  = 
\left( {\bf a}^\dag_{n'l} {\bf a}_{nl} \right) ^{(00)}_{00}  \; ,
\end{equation}
one obtains
\begin{equation}
\label{eq:density_BP}
\langle \Psi_{\alpha J M } \vert  \hat{\delta}(r) \vert \Psi_{\alpha J M } \rangle
= (-1)^{J - M} 
\left( \begin{array}{ccc} J & 0 & J \\ -M  & 0 & M \end{array}    \right)
\langle \Psi_{\alpha  J }  \| \widehat{F}^{(00)0}_{\rho} \| \Psi_{\alpha J } \rangle
\end{equation}
with
\begin{equation}
\label{eq:F_00_LSJ}
\widehat{F}^{(00)0}_{\rho, 0} \; = \;
    - {\sum_{l = 1}} \sqrt{2 \left( 2l+1 \right)}
 \; {\sum_{n,n^{\prime}}}
    \left( {\bf a}^\dag_{n'l} {\bf a}_{nl} \right) ^{(00)0} _0
\;  I_{\rho} \left( n^{\prime}l, nl \right)  \; ,
\end{equation}
and 
\begin{equation}
\label{eq:I_rho}
I_{\rho}\left(n^{\prime}l, nl \right) (r) \; \equiv \;
R_{n'l} (r) R_{nl} (r) r^2 \; .
\end{equation}
The diagonal reduced matrix element (RME) evanuated with the Breit-Pauli eigenvector (\ref{MCHF_BP_exp}) has the following form
\begin{equation}
\label{eq:density_rme_BP}
\langle \Psi_{\alpha J } \| \widehat{F}^{(00)0}_{\rho} \| \Psi_{\alpha J } \rangle
= \sum_{i,j} a^\ast_i a_j \; 
\langle \Phi (\alpha_i L_i S_i  J )  \| \widehat{F}^{(00)0}_{\rho} \| \Phi (\alpha_j L_j S_j  J ) \rangle
\end{equation}
where the RME in the $(LS)J$ coupled basis reduces to
\begin{equation}
\label{eq:density_rme_CSF}
\langle \Phi (\alpha_i L_i S_i  J M)  \| \widehat{F}^{(00)0}_{\rho} \| \Phi (\alpha_j L_j S_j  J M) \rangle
=
\sqrt{\frac{2J+1}{(2L_i+1)(2S_i+1)}} \;
\langle \Phi (\alpha_i L_i S_i )  \| \widehat{F}^{(00)}_{\rho} \| \Phi (\alpha_j L_j S_j  ) \rangle
 \delta_{L_i,L_j} 
 \delta_{S_i,S_j} 
\end{equation}
and
\begin{equation}
\label{eq:Density_operator_SC}
\widehat{F}^{(00)}_{\rho, 0 0} \; = \;
    - {\sum_{l = 1}} \sqrt{2 \left( 2l+1 \right)}
 \; {\sum_{n,n^{\prime}}}
    \left( {\bf a}^\dag_{n'l} {\bf a}_{nl} \right) ^{(00)}_{00} 
\;  I_{\rho} \left( n^{\prime}l, nl \right)  \; .
\end{equation}

From the analogy of the operator (\ref{eq:Density_operator_SC})   and the non-relativistic one-body Hamiltonian operator (see eq.~(A5) of \cite{Olsetal:95a}),
one observes that the angular coefficients of the radial functions $I_{\rho}\left(n^{\prime}l, nl \right) (r)$ 
are identical to those of the one-electron Hamiltonian radial integrals
$I_{n^{\prime}l, nl}$ ,  as anticipated from McWeeny analysis \cite{Wee:92a}. 
These angular coefficients can be derived by working out the  matrix elements of a one--particle 
scalar operator $\widehat{F}_{\rho}^{(00)}$ between configuration state functions with $u$
open shells, as explicitly derived by Gaigalas {\em et al} \cite{Gaietal:01a} who expressed them as a sum over
one--electron contributions
\begin{equation}
\label{eq:one-a}
\langle \Phi (\alpha L S ) \left \| \widehat{F}^{(0 0)}_{\rho} \right\|
 \Phi (\alpha^{\prime} L S  ) \rangle =
  {\sum_{n_i l_i,n_j l_j}}
\langle \Phi (\alpha L S ) \left \| \widehat{F}_{\rho}
 ( n_i l_i, n_j l_j )
 \right\|  \Phi (\alpha^{\prime} L S  ) \rangle
\end{equation}
where
\begin{eqnarray}
\label{eq:mat-one-body}
\lefteqn{\D
\langle \Phi (\alpha L S ) \left \| \widehat{F}_{\rho}
 ( n_i l_i, n_j l_j )
 \right\|  \Phi (\alpha^{\prime} L S  ) \rangle}
 \nonumber  \\[1ex]
 & & =
 \D ( -1)^{\Delta +1} \sqrt{2(2l_i +1)} \; 
 R\left( \lambda_i, \lambda_j, \Lambda
 ^{bra}, \Lambda ^{ket} \right) \, \delta_{l_i , l_j}  \,
 I_{\rho}\left(n_i l_i, n_j l_j \right) 
  \nonumber  \\[1ex]
 & & \times  \left\{ \delta ( n_i , n_j )
 \left( n_i l_i^{N_i}\;\alpha _i Q_i L_i S_i \left\|
 \left[ a^{\left( q \; \; l_i \; s \right)}_{1/2}
 \times  a^{\left( q \; \; l _i \; s
 \right)} _{-1/2}\right] ^{\left( 0 \; 0 \right) }
 \right\|n_{i} l_{i}^{N_{i}} \;\alpha _i Q_i L_i S_i \right)
 \nonumber \right. \\[1ex]
 & & \left. + (1-\delta ( n_i , n_j ))
 \left( n_{i} l_{i}^{N_i}\,\alpha _i Q_i L_i S_i \left\|
 a^{( q \,l _i \, s)}_{1/2}
 \right\|n_{i} l_i^{N_{i}^{\prime }} \,\alpha _i Q_i L_i S_i \right)
 \nonumber \right. \\[1ex]
 & & \left. \times
 \left( n_{j} l_j^{N_j}\;\alpha _j Q_j L_j S_j \left\|
  a^{( q \, l_j \, s)}_{-1/2}
 \right\|n_{j} l_j^{N_{j}^{\prime }} \,\alpha_j Q_j L_j S_j\right) \right\} \; .
\end{eqnarray}
In this last expression, $\lambda \equiv l$ or $s$,
$\langle \Phi (\alpha L S ) \vert $ and $ \vert \Phi (\alpha^{\prime} L S  ) \rangle $
%$\vert CSF \left( L^{\prime } S^{\prime } \right) )$ 
are respectively bra and ket functions with $u$ open
subshells, \\
$\Lambda^{bra} \equiv
\left( L_iS_i, L_jS_j, L_{i^{\prime }}S_{i^{\prime }},
L_{j^{\prime }} S_{j^{\prime }} \right)^{bra}$
and
$\Lambda^{ket}\equiv
\left( L_iS_i, L_jS_j, L_{i^{\prime }}S_{i^{\prime }},
L_{j^{\prime }}S_{j^{\prime }}\right)^{ket}$
denote the
respective sets of active subshell angular momenta. The operators
$a^{( q \, l s)}_{m_q}$ are second quantization operators
in quasispin space of rank $q = 1/2$. The operator
$a^{( q \, l s)}_{1/2 \; m_l \; m_s} =
a^{\left( l \, s \right) + }_{m_l \; m_s}$
creates
electrons with angular momentum quantum numbers $l,m_l,s,m_s$ and its
conjugate $a^{( q \, l \, s)}_{-1/2 \, m_l \, m_s} =
\tilde{a}_{m_l m_s }^{( l \, s )}  =
( -1)^{l+s-m_l-m_s }a_{-m_l \, m_s }^{( l \, s ) }$ annihilates electrons 
with the same quantum numbers  $l,m_l,s,m_s$ in a
given subshell.
The coefficient 
$R\left( \lambda_i, \lambda_j, \Lambda ^{bra}, \Lambda ^{ket} \right)$
is the recoupling matrices in $l$- and $s$- spaces and $\Delta$ is 
a phase factor.

\section{Density matrix and natural orbitals}
Using (\ref{eq:density_BP}), (\ref{eq:density_rme_BP}), (\ref{eq:density_rme_CSF}) and (\ref{eq:mat-one-body}), 
the radial distribution function gets the following form
\begin{equation}
D(r) = r^2 \rho(r)  = \sum_{i j} a^\ast_i D_{i j} (r) a_j 
= \sum_{i j} a^\ast_i \left[ \sum_{l} \sum_{n' n} v_{n n' l}^{i j} I_\rho (n'l,nl) \right] a_j \; ,
\end{equation}
which can be rewritten in a compact form
\begin{equation}
\label{rad_dens_fct_dens_mat}
D(r) =  \sum_{l} \sum_{n' n} \rho^l_{n' n}  I_\rho (n'l,nl)  \; ,
\end{equation}
with
\begin{equation}
\label{rho_l_nn'}
\rho^l_{n' n} = \sum_{i j} a^\ast_i \; v_{n n' l}^{i j} \; a_j \; .
\end{equation}
The $\delta_{l_i , l_j}$ Kronecker  appearing in  (\ref{eq:mat-one-body}) assures the block-structure of the density matrix 
{\boldmath $ \rho$} whose elements are defined by (\ref{rho_l_nn'}) for the $l$-angular symmetry.

The natural orbitals (NO)  are defined as the one-electron functions that diagonalize the 
density matrix {\boldmath $ \rho$}
\begin{equation}
\label{diag_density}
%\mbox{\boldmath $ \tilde{\rho}$}^l =  {\bf C}^l ^{\dagger} \mbox{\boldmath $ \rho$}^l \; {\bf C}^l
{\bf C} ^{\dagger} \mbox{\boldmath $ \rho$} \; {\bf C}
= \mbox{\boldmath $ \tilde{\rho}$} \; .
\end{equation}
Within a specific angular $l$-symmetry, the eigenvalue problem for the relevant $l$-block
\begin{equation}
\label{}
\mbox{\boldmath $ \rho$}^l {\bf C}^l = {\bf C}^l \mbox{\boldmath $ \tilde{\rho}$}^l
%\mbox{\boldmath $ \rho$}^l {\bf C}^l = {\bf C}^l  \mbox{\boldmath $ \lambda$}^l
\end{equation}
defines the natural radial orbitals through the following transformation
\begin{equation}
\tilde{R}_{ k l } (r) = \sum_n c^l_{n,k}  R_{nl} (r) \; .
\end{equation}
The eigenvalues $\{ \lambda^l_k  = \tilde{\rho}^l_{ k k }  \} $ are interpreted as the occupation numbers of the NOs $\{ \tilde{R}_{ k l } (r) \}$.

\section{Algorithm description}

To calculate the radial density function and the natural orbitals from an arbitrary $N$-electron wavefunction $\Psi_{\alpha J M}$, we wrote a FORTRAN implementation of equation (\ref{eq:density_BP}), as an extension of the {\sc atsp2K} package.
The essential part in the calculation of the density function, is the evaluation of the reduced matrix element (\ref{eq:density_rme_BP}).
In pseudo-code, the reduced matrix element (\ref{eq:density_rme_BP}) is written as
%%%%%%%%%%%
\begin{multline}
\label{pseudo_code}
\bra< \Psi_{\alpha J } | 
\red| \widehat{F}^{(00)0}_{\rho} |
\ket| \Psi_{\alpha J } >
= 
\sum_i 
\sum_j
a_i a_j
\sum_{\mu} \sum_{\nu}
I_\rho(\mu,\nu)  \;
\mathbf{UNITELEMENT}(\mu,\nu)
\\
\mathbf{SPIN\_ANGULAR\_DENSITY}(\mathbf{CSF}_i,\mu;\mathbf{CSF}_j,\nu;00)
\,.
\end{multline}
%%%%%%%%%%	
where 
%%%%%%%%%%%
\begin{equation}
\label{eq:26}
\mathbf{UNITELEMENT}(\mu,\nu) =
-[l_\mu,s_\mu]^\half
\delta(l_\mu,l_\nu)
\,.
\end{equation}
%%%%%%%%%%	
The routine SPIN\_ANGULAR\_DENSITY, is inspired by the routine NONHIPER of the {\tt hfs} hyperfine structures program of {\sc ATSP2K}. It organizes the calculation of the spin-angular part of (\ref{eq:mat-one-body}) 
by calling the subroutine ONEPARTICLE1 or ONEPARTICLE2 from \cite{Froetal:07a}. ONEPARTICLE1 performs the calculation of the spin-angular part when the one-electron operator acts on one open shell and ONEPARTICLE2 performs the calculation when the operator acts on two open shells. Both calculate the spin-angular part using the expresion (\ref{eq:mat-one-body}) 
in which $ I_{\rho}  ( n_i l_i, n_j l_j  ) = 1$. The products of the weight factors with the corresponding spin-angular 
part are stored and accumulated in the two dimensional array FACTORMATRIX$(\mu, \nu)$ where the rows and columns 
are defined by the $(n l)$ subshell quantum numbers of the bra and ket, respectively. FACTORMATRIX is the precursor 
of the density matrix~(\ref{rho_l_nn'}).
The products of the array elements with their corresponding radial part 
$I_{\rho}\left(n^{\prime}l, nl \right)$  are accumulated to build the radial distribution 
function~(\ref{rad_dens_fct_dens_mat}). 
The reader is referred to the flowchart in figure \ref{fig_flowchart} for a schematic overview of the calculation of the density function.

The NOs are obtained by diagonalizing this matrix and using the eigenvectors to construct the orbitals. The diagonalization of the density matrix (\ref{diag_density}) is performed using the {\tt DSYEV} subroutine from the Lapack~\cite{LAPACK} library. This routine computes all eigenvalues and eigenvectors for a given real symmetric matrix. The NOs are ordered and labelled according to their occupation numbers $\{ \lambda^l_k  = \tilde{\rho}^l_{ k k }  \} $.

Most of the subroutines needed for {\tt density} exist in {\tt hfs} of {\sc ATSP2K}, besides the routines from the  {\sc ATSP2K} libraries. 
The new  modules are 
{\tt density.f}, {\tt spin\_angular\_density.f} and  {\tt unitelement.f}.
The code {\tt readwfn.f} that reads in the wave functions differs from the one encountered in {\tt hfs} by the
 {\tt COMMON/ADATA2/AT,TT,ELNAME(NWD)}  needed to store the ATOM, TERM and ELNAME variables.

As an illustration, an interactive session is described in appendix~\ref{ap_intses}, for a $n=3$ CAS-MCHF expansion of the beryllium ground state (63~CSFs). 
Upon execution of {\tt density}, the user is asked to specify the name of the data files, which were obtained from an {\sc ATSP2K} run. {\tt density} then reads  the CSF weights ($\{ c_i \}$ and $\{ a_i \}$ for the non-relativisic and Breit-Pauli expansions, respectively), the configuration state functions quantum numbers and the radial functions from the files. 
The conventions of the data and the file types, summarized in table~\ref{file_convention}, were adopted from {\sc ATSP2K}.  However, for a relativistic calculation, the {\tt .j} should be renamed file {\tt.l} and edited to extract the selected relativistic $J$-eigenvector of interest.

In an interactive session, {\tt density} asks the user a few questions concerning the output and wether the NOs should be evaluated. In table~\ref{Questions} we list and comment the questions. Most of the output, however, is written to disk. The output files produced by the program  are summarized in table~\ref{output_file_convention}. The {\tt name.d} file, which is always generated, contains the radial distribution  $D(r)$ and density  $\rho(r)$ functions.
The {\tt density} program by default generates some output to the standard out: the ``modified electron density''~\cite{Auf:82a} at the nucleus ($\rho(0) = 4 \pi \rho({\bf 0}) $), the occupation numbers of the natural orbitals, with their composition in terms of the original orbitals, and as a final check, the integral of the density function that should give the total number of electrons according to~(\ref{integral_density}).
The $\overline{P}_{nl}(\rho)=  r^{-1/2} P_{nl}(r)$ functions appearing in the files {\tt name.plt} and {\tt name.n} are defined in the logaritmic variable $\rho = \log_e (Zr)$
\cite{Fro:77a} for the original and natural orbitals respectively. 
If the user asks for more details ({\tt `yes'} to the question {\tt PRINT ALL DATA (y/*))}, {\tt density} prints out the contributions to the reduced matrix element~(\ref{pseudo_code}), providing for each pair $(i,j)$ of CSFs, the labels $(\mu, \nu)$ of the orbitals involved, the corresponding spin-angular coefficient, together with the relevant weights product $(a_i a_j)$. Using this option, the user also gets the contributions to the modified density at the nucleus, the norm of the input orbitals, the matrix elements of the density matrix, and the natural orbitals (before they are sorted according to their occupation number), with their complete eigenvector composition. 

To install the program (FORTRAN 90 compilation and linking with the {\sc ATSP2K} libraries), the provided {\sc Install}  script  should be edited 
to set the appropriate path and environment variables.

\section{Applications and examples}

To illustrate the data in the output files, we plotted in figure \ref{fig_Be} the radial density distribution $D(r) =  r^2 \rho(r) $ from the {\tt .d} output file calculated for  a CAS-MCHF wave function of the beryllium ground state (Be $1s^2 2s^2 \; ^1S$), using a $n=9$ orbital active set. 
In the same figure, the Hartree-Fock radial density is compared with the one obtained with two correlation models: i) the $n=2$ CAS-MCHF expansion, largely dominated by the near-degeneracy mixing associated to the Layzer complex $1s^2 \{2s^2 + 2p^2 \}$ and ii) the $n=9$ CAS-MCHF. From the plotted results we notice that  the density of the $n=2$ calculation already contains the major correlation effects, compared to the $n=9$ calculation. Indeed, the density does not seem to change a lot by going from the $n=2$ to the $n=9$ orbital basis, the valence double excitation $1s^2 2p^2$ contributing for 9.7\% of the wave function.  From the energy point of view however, this observation is somewhat surprising (see table~\ref{table_BE}): the correlation energy associated to the $n=2$ CAS-MCHF solution ``only'' represents  47\% of the $n=9$ correlation energy.
%This could have been anticipated since from an energetic point of view the $n=2$ already contains 
% a large part of the correlation energy energy, as can be seen from the values reported in table . 

In a separated pair-MCHF approach, the reduced forms of the CSF expansions are often used  to get a compact multiconfiguration representation of the state and  to avoid possible variational redundancies between orbital rotations and mixing coefficients transformations. For some specific cases, the so-produced MCHF one-electron functions are 
nothing else than the natural orbitals~\cite{Fro:77a}. For expansions closed under orbital rotations, one can test our density computational tool by: 1) perfoming an  (unreduced) MCHF calculation, 2) obtain the natural orbitals from the diagonalization of the density matrix and 3) making a CI calculation in the resulting NO basis. Both calculations should yield  the same total energy for two rather different representations of the same total wave function. Amongst the two, the NO-CSF expansion is naturally condensed. This is illustrated in table \ref{table_BE_NO_CSF} for a $n=5$ SD-MCHF valence correlation calculation on the ground state of Be ($E =  -14.619~083$~a.u., using a Hartree-Fock frozen core).
 The eigenvectors calculated in both MCHF and NO one-electron bases  are reported and compared to each other. Note that, in this specific case (a pair of $\; ^1S^{e}$ symmetry), the transformation that diagonalizes the density matrix eliminates the off-diagonal ($n \neq n' $) contributions $1s^2 nl n'l$~\cite{Fro:73a}. The reduction in the number of CSFs ($30 \rightarrow 15$) through the use of NOs is quite impressive. 
For a $n=6$ SD-MCHF valence correlation calculation the CI-NO approach yields a CSF expansion with 29 terms less and for a CAS-MCHF $n=9$ wave function (271~733 CSFs), the NO basis leads to a reduction of 15~695 CSFs. 
%  {\tt 222268 - 206573 = 15695} (number of zeros in NO basis - number of zeros in original basis : number of new zeros

As a third example,  we illustrate the influence of relativistic effects -- in the Breit-Pauli approximation -- on the density function of the Be-like O$^{+4}$ atom, by comparing the densities of the fine-structure states $1s^2 2s 2p \; ^3P^\circ_0$, $^3P^\circ_1$ and $^3P^\circ_2$. From the plots in figure \ref{fig_BP} and the data given in table \ref{table_BP} we observe that the largest energy difference corresponds to the largest difference in density function. More bound is the level, higher is the electron density in the inner region.
% check the total energies

When studying the electron affinities, it is often interesting to investigate the differential correlation effects between the negative ion and the neutral system \cite{GodFro:99b}. Figure~\ref{fig_S} displays the density functions $D(r)$ of both the [Ne]$3s^2 3p^4 \; ^3P$ ground state of neutral Sulphur (S) and the [Ne]$3s^2 3p^5 \; ^2P^\circ$ ground state of the negative ion S$^-$, evaluated with elaborate correlation models \cite{Caretal:09a}, together with their difference $\Delta D(r)$. The latter  reveals where the ``extra'' electron lies and its integration gives one, as it should.

\section*{Ackowledgements}

The authors acknowledge Thomas Carette, Paul Geerlings and Brian Sutcliffe for helpful discussions. M.~Godefroid thanks the Communaut\'e fran\c{c}aise of Belgium (Action de Recherche Concert\'ee)
and the Belgian National Fund for Scientific Research (FRFC/IISN Convention) for financial support.

%%%%%%%%%%%%%%%%%%%%%%%

%%%%%%%%%%%%%%%%%%%%%%%%%%%%%%%%%%%%%%%%%%%%%%%%%%%%%%%%%%%%%%%%%%%%%%%%%%%%%%%%
%
% BIBLIOGRAPHY
%
%%%%%%%%%%%%%%%%%%%%%%%%%%%%%%%%%%%%%%%%%%%%%%%%%%%%%%%%%%%%%%%%%%%%%%%%%%%%%%%%
%\bibliographystyle{pccp}
%\bibliography{d:/mrg/atoms}

\appendix
%%%%%%%%%%%%%%%%%%%%%%%%%%%%%%%%%%%%%%%%%%%%%%%%%%%%%%%%%%%%%%%%%%%%%%%%%
%
%  AN INTERACTIVE SESSION
%
%%%%%%%%%%%%%%%%%%%%%%%%%%%%%%%%%%%%%%%%%%%%%%%%%%%%%%%%%%%%%%%%%%%%%%%%%

\section{An interactive session} \label{ap_intses}
%%%%%%%%%%%%%%%%%%%%%%%
\begin{scriptsize}
\begin{verbatim}
$cat n3.c
                                                                
                                                                
  1s( 2)  2s( 2)
 1S0 1S0 1S 
  1s( 2)  2s( 1)  3s( 1)
 1S0 2S1 2S1 2S  1S 
  1s( 2)  2p( 2)
 1S0 1S0 1S 
  1s( 2)  2p( 1)  3p( 1)
 1S0 2P1 2P1 2P  1S 
  1s( 2)  3s( 2)
 1S0 1S0 1S 
  1s( 2)  3p( 2)
 1S0 1S0 1S 
  1s( 2)  3d( 2)
 1S0 1S0 1S 
  1s( 1)  2s( 2)  3s( 1)
...
  3p( 4)
 1S0
  3p( 2)  3d( 2)
 1S0 1S0 1S 
  3p( 2)  3d( 2)
 1D2 1D2 1S 
  3p( 2)  3d( 2)
 3P2 3P2 1S 
  3d( 4)
 1S0
  3d( 4)
 1S4
*

$cat n3.l
  Be      Z =   4.0  NEL =   0   NCFG =     63


  2*J =    0  NUMBER =   1
     Ssms =      0.484179758
     1   -14.654414586  1s(2).2s(2)_1S
 0.95181933 0.00029779 0.30027819 0.00037936-0.00118903-0.00023749-0.01763502
 0.00019391-0.04365498 0.00316223-0.00663489 0.00377678-0.00043175 0.00173694
-0.00032836-0.00086628 0.00002903 0.00159570-0.00108079-0.00181964 0.00002085
-0.00001498-0.00000941 0.00415462 0.00703251-0.02349037 0.02848932-0.00012045
-0.00265663 0.00007304-0.00016224-0.00019179 0.00009463 0.00016368 0.00001426
 0.00011178-0.00003997 0.00061770 0.00194285-0.00725367-0.00004543 0.00897260
 0.00000543-0.00001468-0.00005233-0.00000271-0.00000675-0.00000998-0.00002343
 0.00003445 0.00000814-0.00013590-0.00000014-0.00000679-0.00002475 0.00042840
 0.00000708-0.00000844-0.00052396 0.00000473 0.00000004 0.00000119 0.00000002
 
 $density 
 Density calculation, Summer 2009 
 Give <name> of the <name>.c, <name>.l <name>.w files:
n3
 Files: n3 
 
 
 PRINT THE ORBITALS  (*/n) 
 
 Printout orbitals

 PRINT THE MATRIX (*/n) 
 
 Printout the matrix
 
 CALCULATE NATURAL ORBITALS (*/n) 
 
 Calculate natural orbitals
 
 PRINT ALL DATA (y/*)
 
 Do not print all informations

 ANALYSING THE CALCULATION
 =========================

 ACCURACY IS SET TO    1.0000000000000007E-016


 STATE  (WITH      63 CONFIGURATIONS):
 ------------------------------------


 THERE ARE  6 ORBITALS AS FOLLOWS:

       1s  2s  2p  3s  3p  3d

 THERE ARE  0 CLOSED SUBSHELLS COMMON TO ALL CONFIGURATIONS AS FOLLOWS:


 
 NORM OF WEIGHTS =    1.000000004562740     
 
 ATOM Be     TERM 1Se   
 
 ALL WAVEFUNCTIONS EXIST.
 
 
 START OF THE DENSITY CALCULATION
 ================================
 
 
 MODIFIED ELECTRON DENSITY AT THE NUCLEUS:
 O =      444.31734212383130000

 EIGENVECTOR:

  1 = Eigenvalue  6 :   0.19968595313710157E+01
 1s '=
 -0.99450714610441153E+00      1s     AZ=  0.14887071657598840E+02
  0.10466865508139691E+00      2s     AZ=  0.10194455194727872E+01
 -0.94819357413400507E-04      3s     AZ=  0.23772474518338814E+02

  2 = Eigenvalue  5 :   0.18147702141513149E+01
 2s '=
  0.99450712354440736E+00      2s     AZ=  0.10194455194727872E+01
  0.10466862950576727E+00      1s     AZ=  0.14887071657598840E+02
  0.24334504925053317E-03      3s     AZ=  0.23772474518338814E+02

  3 = Eigenvalue  2 :   0.12503444827533565E-02
 3s '=
  0.99999996589623770E+00      3s     AZ=  0.23772474518338814E+02
 -0.11976912542329965E-03      1s     AZ=  0.14887071657598840E+02
 -0.23208377825771107E-03      2s     AZ=  0.10194455194727872E+01

  4 = Eigenvalue  4 :   0.18458626963217419E+00
 2p '=
 -0.99999853877956664E+00      2p     AZ=  0.15057313981228271E+01
 -0.17095141798808027E-02      3p     AZ=  0.51881186928943286E+02

  5 = Eigenvalue  3 :   0.18993473382529018E-02
 3p '=
 -0.99999853877956664E+00      3p     AZ=  0.51881186928943286E+02
  0.17095141798808027E-02      2p     AZ=  0.15057313981228271E+01

  6 = Eigenvalue  1 :   0.63431127544789989E-03
 3d '=
  0.10000000000000000E+01      3d     AZ=  0.31738718272621771E+00

 SUM OF EIGENVALUES     4.000000018250959     
 
 INTEGRAL OF THE DENSITY FUNCTION:
 N =        4.00000001825096200
 
 DENSITY FUNCTION IS IN FILE n3.d                    
 END.

\end{verbatim}
\end{scriptsize}

\newpage

\begin{table}
\begin{center}
\begin{tabular}{l l} 
{\sl extension} & {\sl data in the file}\\
\hline
{\tt .c} &  configuration state function (CSF) expansion \\
{\tt .w} &  radial wave functions (numerical values in binary form)\\
{\tt .l} &  expansion coefficients from a non-relativistic ($LS$) calculation\\
{\tt .j} &  expansion coefficients from a Breit-Pauli ($LSJ$) calculation \\
%{\tt .t} &  term dependence of a .j file\\
\hline
\end{tabular}
\end{center}
\caption{File convention}
\label{file_convention}
\end{table}

\begin{table}
\begin{center}
\begin{tabular}{lclc|} 
Question & Answer & Implication\\
\hline
{\tt PRINT THE ORBITALS  (*/n)} &  y &  The input radial functions  will be written to {\tt .plt} . \\
% & n &  $R_{nl}(r)$ will not be written. \\
{\tt PRINT THE MATRIX (*/n)} &  y & The density matrix will be written to {\tt .matrix} .\\
% & n & The density matrix will not be written. \\
{\tt CALCULATE NATURAL ORBITALS (*/n) } & y & Calculate the NOs and write them on {\tt .n} (formatted) \\
& & and {\tt .nw} (unformatted) files. \\
% & n & Don't calculate the NOs \\
{\tt PRINT ALL DATA (y/*)} & y & Detailed output written to std out: \\
& &  {\tt MODIFIED DENSITY AT THE NUCLEUS}  \\ %?
& &  {\tt NORM OF THE ORBITALS} \\
& &  {\tt DENSITY MATRIX} \\
& &  {\tt EIGENVALUES AND EIGENVECTORS} \\
% & n & Don't print extra information to standard out \\
\hline
\end{tabular}
\end{center}
\caption{Questions {\tt density} asks the user. ``{\tt *}" indicates the default answer.}
\label{Questions}
\end{table}

\begin{table}
\begin{center}
\begin{tabular}{l l} 
{\sl extension} & {\sl data in the file}\\
\hline
{\tt .plt} &   $r_i$, $R_{nl}(r_i)$, $P_{nl} (r_i)=r_i R_{nl}(r_i)$, $\overline{P}_{nl}(\rho_i)=  r_i^{-1/2} P_{nl}(r_i)$ \\
{\tt .d} &  $r_i$, $ \rho(r_i)$, $D(r_i) =  r_i^2 \rho(r_i) $\\
{\tt .n} &  $r_i$, $\tilde{R}_{nl}(r_i)$, $\tilde{P}_{nl} (r_i)=r_i \tilde{R}_{nl}(r_i)$, $\overline{\tilde{P}}_{nl}(\rho_i)$ for Natural Orbitals\\
{\tt .nw} &  analogue of {\tt .w} for the Natural Orbitals (contains $\overline{\tilde{P}}_{nl}(\rho_i)$) \\
\hline
\end{tabular}
\end{center}
\caption{Output files created by {\tt density}}
\label{output_file_convention}
\end{table}

\begin{table}
\begin{center}
\begin{tabular}{ccc}
model & energy (a.u.) & correlation energy (a.u.) \\
\hline
HF  &  -14.573~023 & \\ % 14.573023159
$n=2$-CAS & -14.616~856 & $E^{n=2} - E^{HF} = 0.043~832$\\ % 14.616855562 0.0438324
$n=9$-CAS & -14.667~013 & $E^{n=9} - E^{HF} = 0.093~986$\\ %14.667012750 0.09398959
 \hline
\end{tabular}
\end{center}
\caption{Total energy for the ground state of Be with different correlation models.}
\label{table_BE}
\end{table}

\begin{table}
%\begin{footnotesize}
\begin{scriptsize}
\begin{center}
\begin{tabular}{lrr}
\multicolumn{1}{c}{CSF}  &  \multicolumn{1}{c}{MCHF basis}  &  \multicolumn{1}{c}{Natural orbital basis} \\
\hline 
%%%%%%%%%%%%%%%%%%%%%%%%
{\tt  1s( 2)  2s( 2) } &  0.95282855 &  0.95370264 \\
{\tt 1S0 1S0 1S} &  & \\
{\tt  1s( 2)  2s( 1)  3s( 1) } &  0.03858929 &  0.00000000 \\
{\tt 1S0 2S1 2S1 2S  1S} & & \\
{\tt  1s( 2)  2s( 1)  4s( 1)} & -0.01524193 &  0.00000000 \\
{\tt 1S0 2S1 2S1 2S  1S} & & \\
{\tt  1s( 2)  2s( 1)  5s( 1)} &  0.00133387 & -0.00000001 \\
{\tt 1S0 2S1 2S1 2S  1S} & & \\
{\tt  1s( 2)  2p( 2)} &  0.00133387 & 0.29736974 \\
{\tt 1S0 1S0 1S} & & \\
{\tt  1s( 2)  2p( 1)  3p( 1)} & -0.00032489 &  0.00000000 \\
{\tt 1S0 2P1 2P1 2P  1S} & & \\
{\tt  1s( 2)  2p( 1)  4p( 1)} & -0.00019862 &  0.00000000 \\
{\tt 1S0 2P1 2P1 2P  1S} & & \\
{\tt  1s( 2)  2p( 1)  5p( 1)} &  0.00089172 & -0.00000001 \\
{\tt 1S0 2P1 2P1 2P  1S} & & \\
{\tt  1s( 2)  3s( 2)} & -0.03930620 & -0.04031077 \\
{\tt 1S0 1S0 1S} & & \\
{\tt  1s( 2)  3s( 1)  4s( 1)} & -0.00463218 &  0.00000000 \\
{\tt 1S0 2S1 2S1 2S  1S} & & \\
{\tt  1s( 2)  3s( 1)  5s( 1)} &  0.00091733 &  0.00000000 \\
{\tt 1S0 2S1 2S1 2S  1S} & & \\
{\tt  1s( 2)  3p( 2)} &  0.29736945 &  0.00532117 \\
{\tt 1S0 1S0 1S} & & \\
{\tt  1s( 2)  3p( 1)  4p( 1)} & -0.00003969 &   0.00000000 \\
{\tt 1S0 2P1 2P1 2P  1S} & & \\
{\tt  1s( 2)  3p( 1)  5p( 1)} &  0.00024549 &  0.00000000 \\
{\tt 1S0 2P1 2P1 2P  1S} & & \\
{\tt  1s( 2)  3d( 2)} & -0.01669194 & -0.01669247  \\
{\tt 1S0 1S0 1S} & & \\
{\tt  1s( 2)  3d( 1)  4d( 1)} &  0.00005217 &  0.00000000 \\
{\tt 1S0 2D1 2D1 2D  1S} & & \\
{\tt  1s( 2)  3d( 1)  5d( 1)} & -0.00011379 &  0.00000000 \\
{\tt 1S0 2D1 2D1 2D  1S} & & \\
{\tt  1s( 2)  4s( 2)} & -0.00422002 & -0.00432946 \\
{\tt 1S0 1S0 1S} & & \\
{\tt  1s( 2)  4s( 1)  5s( 1)} & -0.00107435 &  0.00000000 \\
{\tt 1S0 2S1 2S1 2S  1S} & & \\
{\tt  1s( 2)  4p( 2)} &  0.00182955 &  0.00184355 \\
{\tt 1S0 1S0 1S} & & \\
{\tt  1s( 2)  4p( 1)  5p( 1)} &  0.00032865 &  0.00000000 \\
{\tt 1S0 2P1 2P1 2P  1S} & & \\
{\tt  1s( 2)  4d( 2) } & -0.00361419 &  -0.00363174 \\
{\tt 1S0 1S0 1S} & & \\
{\tt  1s( 2)  4d( 1)  5d( 1)} &   0.00030308 & 0.00000000 \\
{\tt 1S0 2D1 2D1 2D  1S} & & \\
{\tt  1s( 2)  4f( 2)} &  0.00618640 &  0.00621375 \\
{\tt 1S0 1S0 1S} & & \\
{\tt  1s( 2)  4f( 1)  5f( 1)} & -0.00048678 &  0.00000000 \\
{\tt 1S0 2F1 2F1 2F  1S} & & \\
{\tt  1s( 2)  5s( 2)} & -0.00160546 & -0.00136552 \\
{\tt 1S0 1S0 1S} & & \\
{\tt  1s( 2)  5p( 2)} & -0.00141498 & -0.00149216 \\
{\tt 1S0 1S0 1S} & & \\
{\tt  1s( 2)  5d( 2)} & -0.00103723 & -0.00101914 \\
{\tt 1S0 1S0 1S} & & \\
{\tt  1s( 2)  5f( 2)} &  0.00188266 &  0.00185530 \\
{\tt 1S0 1S0 1S} & & \\
{\tt  1s( 2)  5g( 2)} & -0.00284386 & -0.00284386 \\
{\tt 1S0 1S0 1S} & & \\
{\tt *} & & \\
\hline
\end{tabular}
\end{center}
\caption{Comparison of Be $n=5$-valence eigenvectors in the MCHF and NO bases.}
\label{table_BE_NO_CSF}
\end{scriptsize}
\end{table}

\newpage

\begin{table}
\begin{center}
\begin{tabular}{rcc}
model & energy (a.u.) & energy difference (a.u.)\\
\hline
$1s^2 2s2p \; ^3P^\circ_0$ & -68.032~086 & \\ %68.032085517
$^3P^\circ_1$ & -68.031~473 &$\Delta E_{10} = 0.000~613$\\ %68.031472770 0.00061275
$^3P^\circ_2$ & -68.030~102 &$\Delta E_{21} = 0.001~370$\\ %68.030102285 0.00137048
 \hline
\end{tabular}
\end{center}
\caption{Fine structure total energies of O$^{+4}$~$1s^2 2s2p \; ^3P^\circ$}
\label{table_BP}
\end{table}

\begin{figure}[p]
\begin{center} 
\resizebox{90mm}{!}{\includegraphics[clip,angle=0]{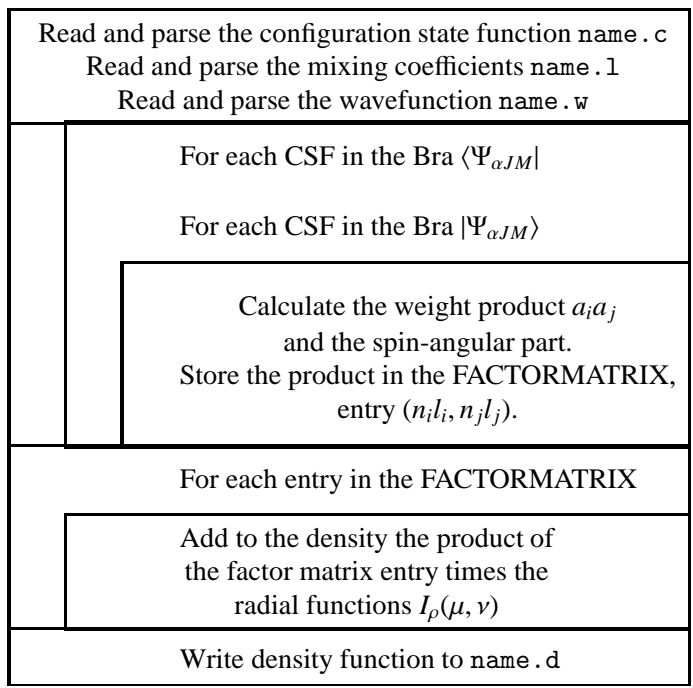}}
\caption{Flowchart of the Density program}
\label{fig_flowchart}
\end{center}
\end{figure}

\begin{figure}[h]
\begin{center}
\psfrag{R(r)}{{\Huge $D(r)$}}
\psfrag{D(r)n=2 - D(r)HFHF}{{\Huge $D(r)^{n=2} - D(r)^{HF}$}}
\psfrag{D(r)n=9 - D(r)HFHF}{{\Huge $D(r)^{n=9} - D(r)^{HF}$}}
\psfrag{r}{{\Huge $r / a_0$}}
%\psfrag{D}{{\Huge $ / a_0^{-1}$}}
\psfrag{D}{{\Huge $ D(r)~\mbox{or}~\Delta D(r)/ a_0^{-1}$}}
\resizebox{90mm}{!}{\includegraphics[clip,angle=-90]{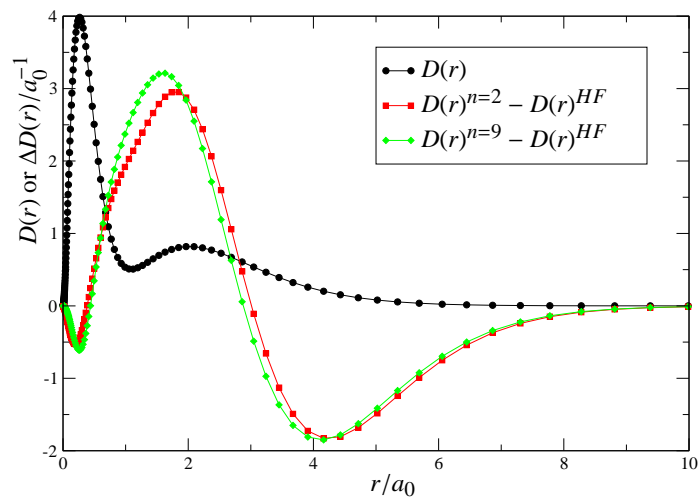}}
\caption{Density of  Be~$\; ^1S$ ground state for different CAS-MCHF wave functions. Density differences have been scaled by a factor 100.}
\label{fig_Be}
\end{center}
\end{figure}

\begin{figure}[h]
\begin{center}
\psfrag{R(r)}{{\Huge $D(r)^{^3P_0}$}}
\psfrag{D(r)3P1 - D(r)3P0}{{\Huge  $D(r)^{^3P_1} - D(r)^{^3P_0}$}}
\psfrag{D(r)3P2 - D(r)3P1}{{\Huge $D(r)^{^3P_2} - D(r)^{^3P_1}$}}
\psfrag{r}{{\Huge $r / a_0$}}
%\psfrag{D}{{\Huge $ / a_0^{-1}$}}
\psfrag{D}{{\Huge $ D(r)~\mbox{or}~\Delta D(r)/ a_0^{-1}$}}
\resizebox{90mm}{!}{\includegraphics[clip,angle=-90]{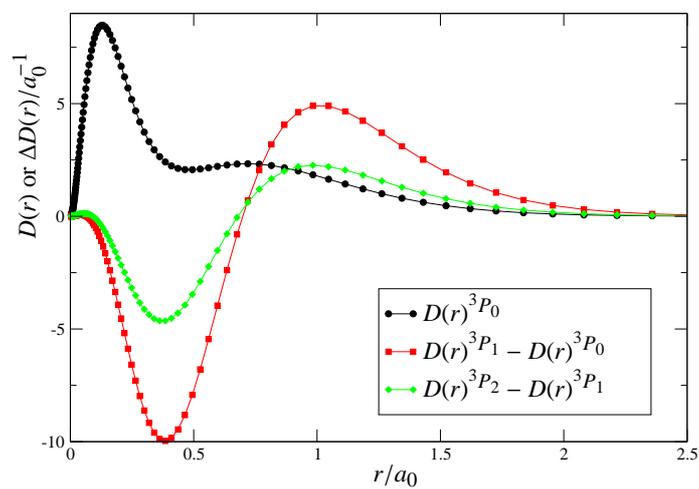}}
\caption{Comparison of the $1s^2 2s 2p \; ^3P^\circ_0$, $^3P^\circ_1$ and $^3P^\circ_2$ radial 
density functions of O$^{+4}$. Density differences have been scaled by a factor 10~000.}
\label{fig_BP}
\end{center}
\end{figure}

\begin{figure}[h]
\begin{center}
\psfrag{D(r)S-}{{\Huge $D(r)^{S^-}$}}
\psfrag{D(r)S}{{\Huge $D(r)^{S}$}}
\psfrag{D(r)S- - D(r)SSSS}{{\Huge $D(r)^{S^-} - D(r)^{S}$}}
\psfrag{r}{{\Huge $r / a_0$}}
%\psfrag{D}{{\Huge $ / a_0^{-1}$}}
\psfrag{D}{{\Huge $ D(r)~\mbox{or}~\Delta D(r)/ a_0^{-1}$}}
\resizebox{90mm}{!}{\includegraphics[clip,angle=-90]{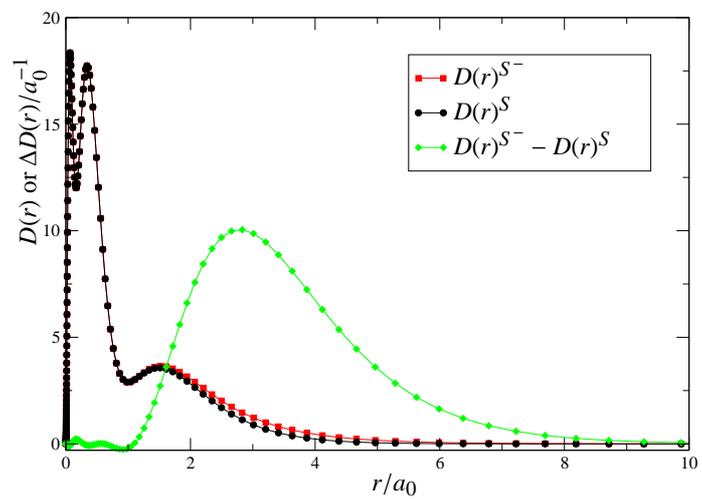}}
\caption{Ground state S and S$^-$ density functions~\cite{Caretal:09a}. Density differences have been scaled by a factor 30.}
\label{fig_S}
\end{center}
\end{figure}

\end{document}